\documentclass[12pt,aps,prb,preprint]{revtex4-1}  
\usepackage{amsmath} 
\usepackage{amsfonts}
\usepackage{amssymb}
\usepackage{graphicx}
\usepackage{subfigure}
\usepackage{epsfig}
\usepackage{hyperref}
\usepackage{bm}

\newcommand{\rmd}{\text{d}}

\begin{document}

\title{Brownian motion in typical microparticle systems}

\author{Silvana Palacios}
\email{corresponding author: silvanapalacios@ciencias.unam.mx}

\author{Victor Romero-Rochin}

\author{Karen Volke-Sepulveda}

\affiliation{Instituto de F\'isica, Universidad Nacional Autonoma de M\'exico,
Apartado Postal 20-364, 01000 Mexico, D.F. Mexico}

\date{\today}

\begin{abstract}
 
Many studies on microscopic systems deal with Brownian particles embedded in media whose densities are different from that of the particles, causing them either to sink or float. The proximity to a wall modifies the friction force the particle experiences, being, instead of the Stokes force assumed by Einstein and Langevin, a function of the separation distance, as described by Faxen. In this work we present a thorough analysis of bidimensional Brownian motion using monodisperse suspensions of latex beads  with controlled temperatures. By means of Langevin's model and Faxen's correction, we calculate the mean separation distance between the spheres and the bottom of the confining vessel.

\end{abstract}

\maketitle

\section{Introduction}
In 1827, Brown observed organic and inorganic microparticles having an uninterrupted and haphazard motion in fluid media.\cite{EINSTEIN} With the same result for many materials, he concluded that all matter should be constituted by a ``primitive molecule''. For 50 years the new phenomenon, later known as Brownian motion, did not received any contribution to its explanation in physical nor mathematical terms. In 1877 Wiener demonstrated that the motion could not be due to convection currents and Dancer discarded electrostatic effects as the reason.\cite{SEGRE}  It was Desaulx and Carbonnelle  who suggested that Brownian motion should constitute a physical phenomenon \textit{per se}, where molecules of the medium hit the microparticle causing it to move randomly. And in 1888 Gouy demonstrated it qualitatively,  by showing experimentally that Brownian motion is independent of the nature of both, the medium and the particle, although  it is certainly affected by temperature.\cite{PERRIN}\\

In his search for evidence of the same idea, the existence of molecules, Einstein  argued in 1905  that the random motion of microscopic particles must be due to the multiple collisions with the yet much smaller molecules of the medium.\cite{EINSTEIN} He derived his celebrated formula relating the macroscopic diffusion coefficient with the temperature and the viscosity of the fluid, and with Avogadro number $N_A$. Moreover, he went further by proposing a simple experiment to measure the mean square displacement (MSD) of a micron-sized particle and to obtain, in turn, the diffusion coefficient. Some years later, in 1908, Langevin  proposed a version of Newton equation for a Brownian particle in a fluid from which Einstein's result naturally followed. The developments by Einstein and Langevin are now milestones of the general theory of Markovian processes.\cite{LANGEVIN}\\ 

In 1906, inspired by Einstein's proposal, Jean Perrin calculated $N_A$ according to Einstein's expression for the MSD. In his experiments, the trajectories of hundreds of gamboge and mastic microparticles with radii in the range of 0.21-5.5$\mu$m were followed by hand using a \textit{camera lucida}.\cite{PERRIN,ATOMS} The particles were embedded in media with different viscosities, being the largest value up to 125 times larger than the smallest one. From this statistics he obtained a value of $N_A=71.5 \times 10^{22}$ particles per mol (see page 90 of Ref. \cite{PERRIN}), which was consistent with the values he obtained based on different studies, and hence this constituted  the first experimental demonstration of Einstein's prediction. \cite{ATOMS}
 Since that moment, Brownian motion acquired an important presence in many topics of physics, moving us to explore it on its fundamentals. \\

Repeating Perrin's classic experiment with modern means has become a very attractive endeavor,  \cite{HARVARD,EJP,BM} not only because of its historical importance but also because it  is still a reference for understanding macroscopic phenomena where the role of the molecular collisions is relevant. It is very important to mention, however, that most studies on Brownian motion, including Perrin's,  involve particles with densities different from that of the medium, and therefore, they either float or sink. This fact is not considered by Einstein's nor by Langevin's theories. In their models, the only forces are the dissipative one, considered as the Stokes drag for a spherical particle, and the stochastic force due to the molecules of the medium. As we point out and study in the present article, once the particles sink to the bottom of the container, the interaction with the wall gives rise to a correction to the viscosity of the medium that substantially modifies Einstein relation. This correction, well known by the scientific community working on optical micromanipulation, may be taken into account by invoking to Faxen's model.\cite{OPTMANIP,OPTRAP}\\

In this work the ultimate goal is not to find the value of Avogadro number $N_A$, but to reconsider the approximations assumed to describe typical microparticle systems. Our interest is to study Brownian motion taking into account the presence of a boundary. Our scheme is the same as Perrin's, but instead of the \textit{camera lucida} we use a coupled-charged device (CCD) and perform a thorough computer analysis of the data. We record the motion of individual latex microspheres and from the videos we obtain their positions as a function of time, which allows us to calculate the MSD. In order to avoid hydrodynamic and electrostatic interactions among particles, we use very low concentrations of particles within an ionic solution. The paper is organized as follows: Section \ref{sec2} briefly describes the Brownian motion model in a timescale larger than the relaxation time of the system, the diffusive regime. In Section \ref{sec3}, the experimental setup and the data analysis algorithm are described. At the end, $N_A$ is calculated without the effective viscosity corrections by analyzing the MSD as function of time. The comparison of this result with the known standard value, shows the necessity of including a different description of the system. In Section \ref{sec4}, we introduce Faxen's correction and, assuming the standard value of $N_A$, the mean vertical distance of the particles from the bottom is calculated. On the basis of our results, we conclude with some remarks and suggestions for further experiments.\\

 \section{Brownian motion mathematical model \label{sec2}}

In Langevin's approach, the  motion of a  spherical microparticle in thermal equilibrium with its surroundings can be expressed by Newton second law where the total force is the addition of a dissipative term and a time-dependent stochastic force.\cite{LANGEVIN} The friction force is the mechanism by means of which the particle looses heat, while the stochastic force, $\mathbf{f}(t)$, representing the collisions of the molecules of the medium with the particle, returns energy to the particle. One should remark that energy can be transferred just by collisions and rotations since the kinetic energy of the molecules is not enough to produce internal excitations in the particle. As the translational contribution to the kinetic energy is much larger than the rotational contribution,\cite{EINSTEIN,FAXCOR} we neglect the rotational dynamics in the model.   \\

According to Stokes, for a spherical particle falling in an unbounded medium with small Reynolds number, the dissipative force is a drag proportional to the velocity of the particle, with coefficient  $\gamma=3\pi \eta d$, being $\eta$ the viscosity of the medium and $d$ the diameter of the particle. Meanwhile $\mathbf{f}(t)$ is  a white noise  force whose first moment is null and its second moment is $\overline{\mathbf{f}(t)\mathbf{f}(t')} = \alpha  \bm{\delta}(t-t')$, which means that it is completely uncorrelated with itself at different times (this is true for time intervals larger than the correlation time of the immersion medium, inherent to the duration of the collisions between molecules). The coefficient $\alpha = 2 RT\gamma /N_A$ as given by the Fluctuation-Dissipation Theorem, expresses the fact that the particle is in thermal equilibrium with the medium; $T$ is the absolute temperature and $R$ the gas constant. The components $\mathbf{x}$, $\mathbf{y}$ and $\mathbf{z}$ of $\mathbf{f}(t)$ are completely uncorrelated  with each other.\cite{VANKAMPEN} Langevin motion equation then reads:

\begin{equation}
\text{m}\frac{\rmd \mathbf{v}}{\rmd t}=-\gamma \mathbf{v} + \mathbf{f}(t),
\label{eqLang}
\end{equation}
 
 \noindent where $\text{m}$ is the mass of the microparticle and $\mathbf{v}$ its velocity. When the density of the particle, $\rho_p$, is different from  the density of the medium, $\rho_m$, one should include a third term given by:
\begin{equation}
    \mathbf{F} = \frac{1}{6}\pi d^3(\rho_m - \rho_p){\mathbf{g}}, \label{F}
  \end{equation}
  
  \noindent with $\mathbf{g}$ the gravitational acceleration. $\mathbf{F}$ causes in general the particle to remain on the bottom or at the top. When the particle is close to a wall, Faxen found that  the consequence is an increase of the effective viscosity the particle experiences. The drag force, then, is better represented by $-k \gamma \mathbf{v}$, where $k$ is given by:\cite{FAXEN,FAXCOR,FAXCOR2}

  \begin{equation}
k=\left[ {1-\frac{9}{32}\left( \frac{d}{h}\right)+\frac{1}{64}\left( \frac{d}{h}\right)^3-\frac{45}{4096}\left( \frac{d}{h}\right)^4-\frac{1}{512}\left( \frac{d}{h}\right)^5 }\right]^{-1},
\label{eqk}
\end{equation}

  \noindent being $h$  the distance from the center of the particle to the wall. Notice that $k$ tends to 1 as $h$ tends to infinity, recovering the Stokes force.\\
  
  Because of the fluctuations along $\mathbf{z}$-axis, $h$ is a function of time, nevertheless we can analyze just the bidimensional motion on the plane assuming  $h$ as an average value of the vertical position of the particle. In this case the gravitational force needs no longer to be explicit in the motion equation.\\
  
   Therefore, we write the Langevin-Faxen bi-dimensional Brownian motion equation for a particle close to one wall as:
  
  \begin{equation}
\text{m}\frac{\rmd \mathbf{v}_\perp}{\rmd t}=-k\gamma \mathbf{v}_\perp + \mathbf{f}_\perp(t).
\label{eqbrownie}
\end{equation}

The factor $\text{m}/(k\gamma) \equiv \tau _R$ is known as the relaxation time. For $\Delta t  \gg \tau _R$ the  probability of finding the particle with velocity $\mathbf{v}_1$ at $t_1=t_0+\Delta t$ is completely independent of $\mathbf{v}_0$ at $t_0$, so the motion is  in a non-correlated regime. Moreover, once the particle relaxes to equilibrium,  it has a velocity distribution that must be the Maxwell velocity distribution.\cite{VANKAMPEN} For typical Brownian systems  $\tau _R$ is $\sim  10^{-7}  $s, which is much smaller than the time resolution for usual imaging systems. For instance, our CCD takes 30fps, hence, we are able to observe only a random walker.\\

In this regime, the term on the left hand side of Eq.~\eqref{eqbrownie} is negligible when divided by $k\gamma$, then we have for the bidimensional position, $\mathbf{r}$:

\begin{equation}
\mathbf{v}_\perp=\frac{\rmd \mathbf{r}}{\rmd t}=\frac{1}{k\gamma} \mathbf{f}_\perp(t).
\label{dr}
\end{equation} 
 
 Equation (\ref{dr}) can be solved as an integral equation. Taking the time average to the solution and considering that $\overline{ \mathbf{r} }=\mathbf{r}_0$, one finds:
 
 \begin{eqnarray}
 \vert \overline{\mathbf{r}-\mathbf{r}_0}\vert & = & 0, \nonumber \\*
\vert \overline{\mathbf{r}^2}-\mathbf{r}_0^2\vert & = &  \vert \overline{ (\mathbf{r}-\mathbf{r}_0)^2}\vert =2\frac{\alpha}{k^2 \gamma ^2} \Delta t =4\frac{RT}{k^2 \gamma N_A} \Delta t , 
\label{msdk}
\end{eqnarray} 

\noindent because of the statistical properties of $\mathbf{f}(t)$. In the limit $k\rightarrow 1$, Eq.~\eqref{msdk} corresponds to Einstein's solution:
 
 \begin{equation}
\vert \overline{ \mathbf{r}^2}-\mathbf{r}_0^2 \vert =4D t,
\label{eqEinst}
\end{equation}

 \noindent with the diffusion coefficient identified as:
 
 \begin{equation}
D=\frac{RT}{\gamma N_A},
\label{D}
\end{equation} 
 which is the celebrated Einstein relation.\cite{EINSTEIN}

 \section{Experiments \label{sec3}}
 
\begin{figure}[h!]
\centering 
\includegraphics[width=.75\textwidth, height=.38\textheight]{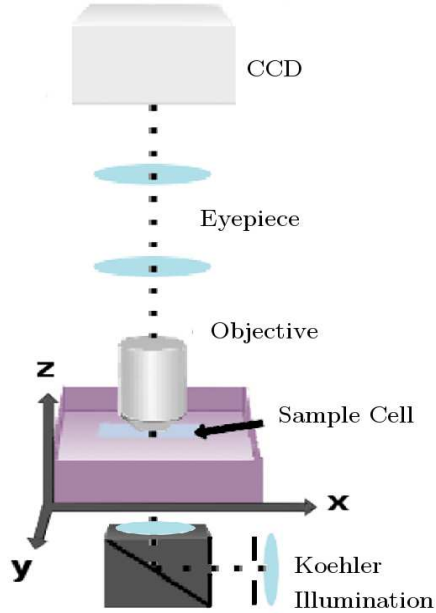}
\caption{Setup for viewing microparticles with Brownian motion.}
\label{setup}
\end{figure}

Figure \ref{setup} depicts the experimental setup for studying Brownian microparticles, which was basically a standard video microscopy system. It consisted of a 100$\times$ immersion objective (N.A. 1.25) and two lenses ($f_\mathrm{L_1}$=50mm and $f_\mathrm{L_2}$=100mm) playing the role of the eyepiece. The total magnification was approximately 140$\times$. As a condenser we used a Koehler illumination system, which provides optimum contrast and resolution, and uniform brightness in the illumination field.\cite{ByW} The video camera was a CCD with a recording speed of 30fps. We restrict our analysis to the motion on the horizontal plane, $\mathbf{x}\mathbf{y}$. \\

Our samples consisted of latex spheres, with either $d_1=1.0\mu$m or  $d_2=2.0\mu$m in diameter (size dispersion$~4\%$)\cite{PART}, suspended in a diluted saline solution. This ionic medium avoids electrostatic interactions among particles.\cite{INTERACCIONES2} The concentrations were low enough to allow the imaging of just one particle per video, avoiding also the hydrodynamic interactions in this way.\cite{INTERACCIONES} The density of the particles, $\rho_p$ =$1.05$g/cm$^3$, is higher than that of the hosting medium, $\rho_m$ =$1.00$g/cm$^3$, and thus they settle on the bottom of the sample cell. The sample cells were shallow cylindrical chambers with 100$\mu$m depth and 0.5cm in diameter made with a dark sticker placed between a microscope slide and a cover slip, both previously cleaned with a free rinsing surfactant (LiquiNox) and with acetone, then sealed with epoxy resin. The samples were placed on a plate with a temperature control system, and the plate was attached to an $\mathbf{x}\mathbf{y}\mathbf{z}$ translational stage, allowing us to choose the particle of interest. It is worth to mention that it was imperative to guarantee the horizontal position of the sample cell, in order to avoid motion currents on the $\mathbf{x}\mathbf{y}$ plane due to the gravitational force. \\

We recorded $Q = 10$ videos for each particle size for over 1 minute, which corresponds to $L =1800$ frames per video. The particles were tracked with a  Matlab code which analyzes frame by frame finding the centroid of the particle image and recording its two-dimensional coordinates as a function of time.  The input parameters for this processing are: 1) the diameter of the object in pixels, 2) the tolerated image noise associated with the change of the image resolution when the particle is out of focus (due to vertical Brownian motion), and 3) the  maximum displacement from one frame to the next one, in order to avoid the tracked particle to be confused with a different one in case many particles are present. 
Since the distances are expressed in pixels, it is necessary to find the equivalence between pixels and micrometers. This was made by measuring in pixels the diameter of the spheres and comparing with the size reported by the manufacturer. From this process  $Q$ data files were generated, these include the two-dimensional positions 
$({\textbf r}_0^{(q)}, {\textbf r}_1^{(q)}, {\textbf r}_2^{(q)}, \dots, {\textbf r}_L^{(q)})$ separated at regular time intervals  $\Delta t = (1/30)$s. Here $q = 1,2, \dots, Q$ denotes the different videos.\\

Since we are assuming a diffusive Markovian stationary process, each step of the particle should be independent of its initial position. We can take advantage of this fact to perform averages with very good statistics by defining  relative displacements as  $\mathbf{\Delta}_l^{(s,q)} = {\textbf r}_{l+s}^{(q)} - {\textbf r}_l^{(q)}$, with $l =0,...,(L-1-s_{max})$ and $s = 1,2, \dots, s_{max}$. Therefore, for every time $t = s \Delta t$ we calculate the MSD as

\begin{equation}
 \overline{({\textbf r}_s - {\textbf r}_0)^2} =  \frac{1}{Q }\sum_{q=1}^Q \left[ \frac{1}{L-1-s_{max}}  \sum_{l=0}^{L-1-s_{max}} \left(\mathbf {\Delta}_l^{(s,q)}\right)^2 \right],
 \end{equation}
 
In our case, we restricted our analysis to the time interval $t\in[0,10)$s, which implies that $s_{max}=300$. Therefore, a total of 15 000 data were used for the calculation of each MSD point, 1500 per video, thus making negligible the statistical error.\cite{SD}\\

\begin{figure}[h!]
\label{primera}
\centering \subfiguretopcaptrue 
\begin{tabular}{cc}
\subfigure[ ]{\label{1}
\includegraphics[width=.49\textwidth, height=.25\textheight]{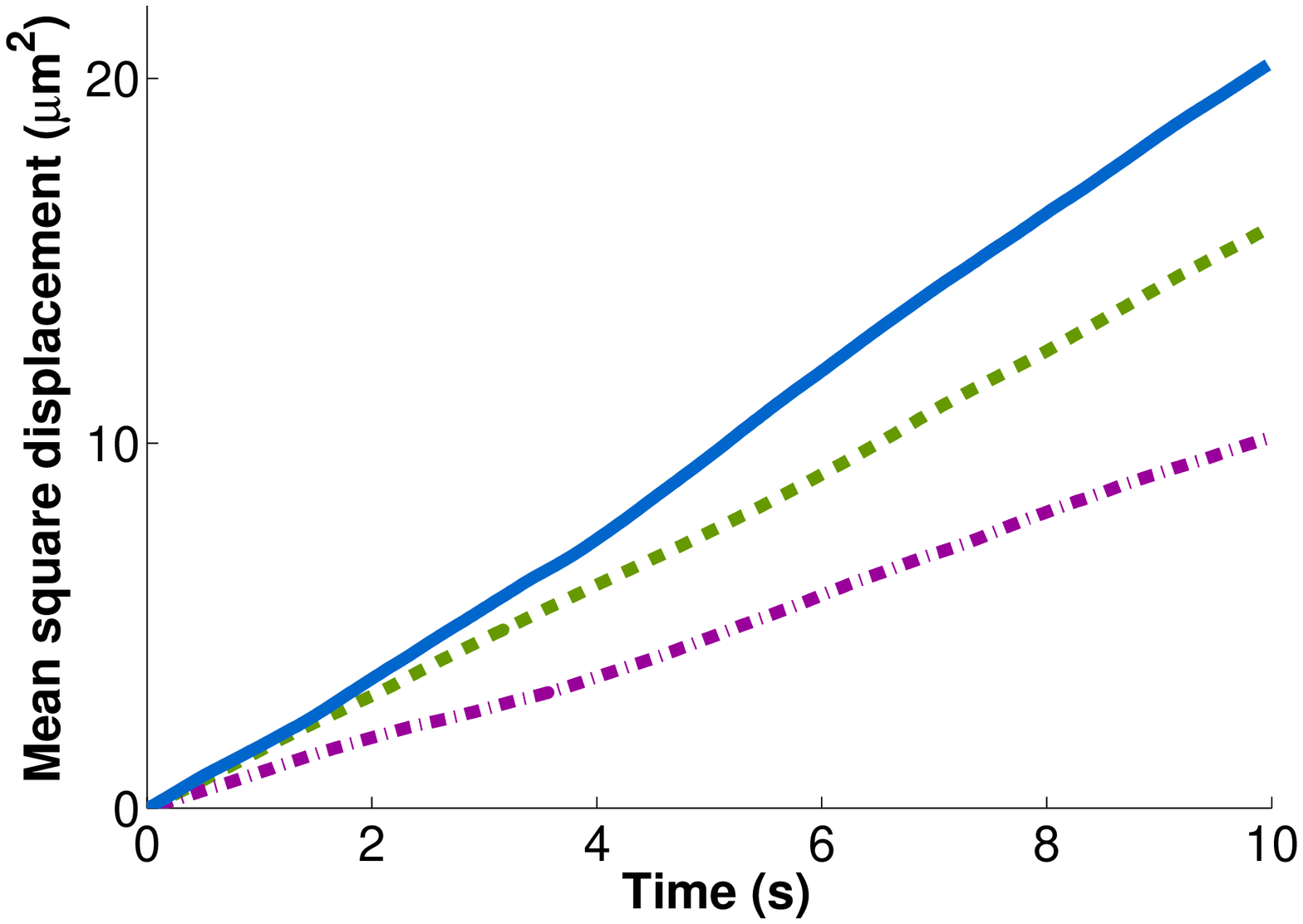}}
\subfigure[ ]{\label{2}
\includegraphics[width=.49\textwidth, height=.25\textheight]{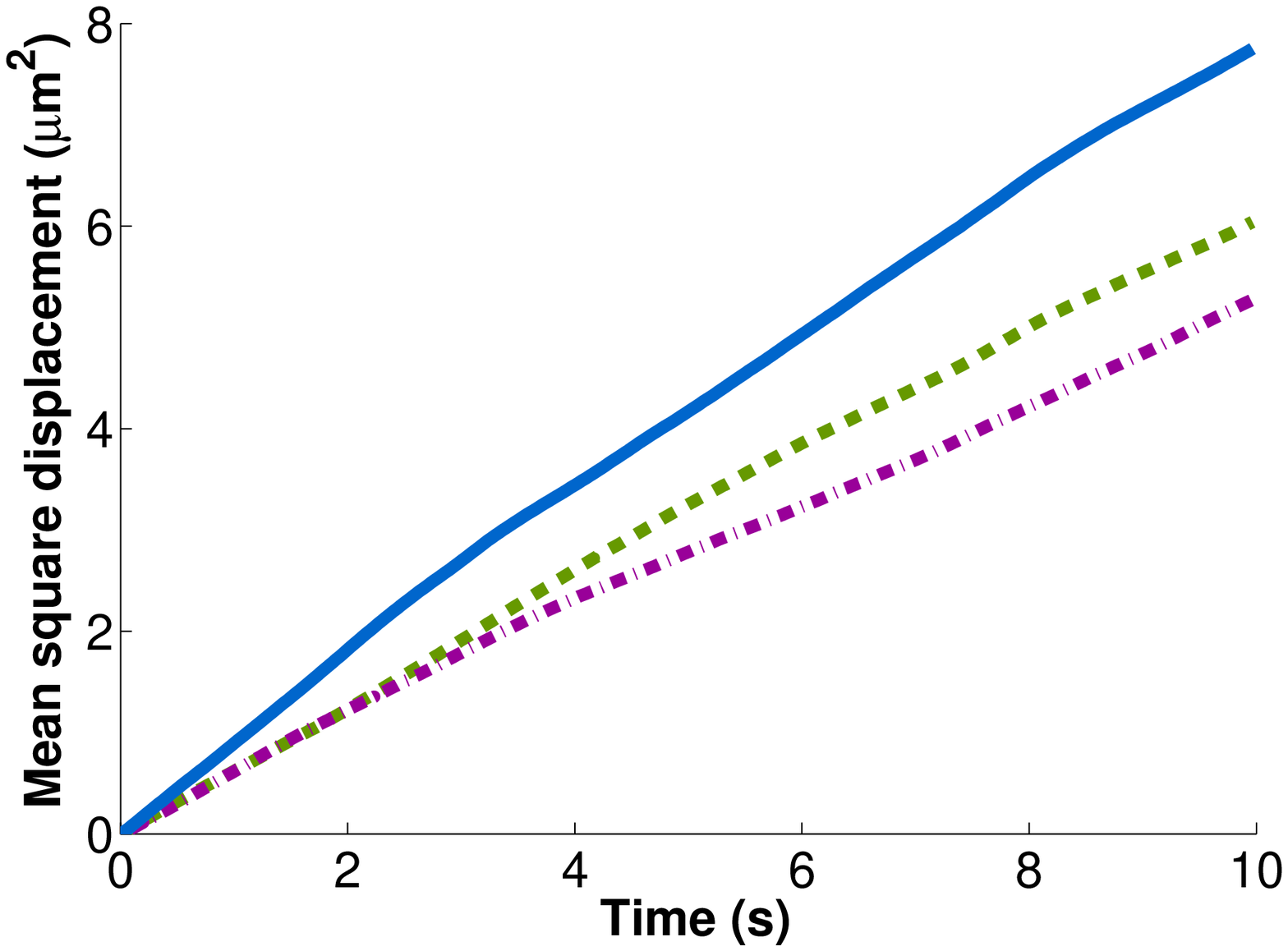}}
\end{tabular}
\caption{(Color online) MSD vs time for latex particles with $d_1=1.0\mu$m (a) and  $d_2=2.0\mu$m  (b) in diameter. Dot-dashed lines correspond to $T=20^\text{o}$C, dashed lines to 40$^\text{o}$C and solid lines to 60$^\text{o}$C.}
\label{graf}
\end{figure}

Figure \ref{graf} shows the plots obtained for the MSD as a function of time for both sizes of particles at different temperatures,  20, 40 and 60$^\text{o}$C. The linear regression through the origin for these functions provided us a value for the diffusion coefficient $D$ according to Eqs.~\eqref{eqEinst} and \eqref{D},  from which we could obtain Avogadro number $N_A$. For instance, for $d_1$ and $T=20^\text{o}$C, we obtained $N_A=9.584 \times 10^{23}$part/mol. Uncertainty on this value can be due just to three factors: 1) reported dispersion size of the spheres 2) measurement in pixels of the image of the particle 3) accuracy of the temperature controller. In any case, they are not enough to explain the large difference with reported values of $N_A$ by different techniques.\cite{AVOGADRO} This incite us to use Eq.~\eqref{msdk} instead of Eq.~\eqref{eqEinst}.\\

 \section{Faxen's boundary conditions \label{sec4}}
 By analyzing Fig. \ref{graf}, we found the linear dependence of MSD with time, nevertheless the proportionality constant is substantially smaller than the Einstein coefficient, $4D$. On the other hand, the use of Eq.~\eqref{eqbrownie} makes possible either to determinate $N_A$ more accurately by knowing the average vertical position of the particle, or to determinate the average vertical position of the particle, $h$,  by assuming the standard $N_A$ value. We explored the second option.\\
 
\begin{figure}[h!]
\centering 
\includegraphics[width=.56\textwidth, height=.3\textheight]{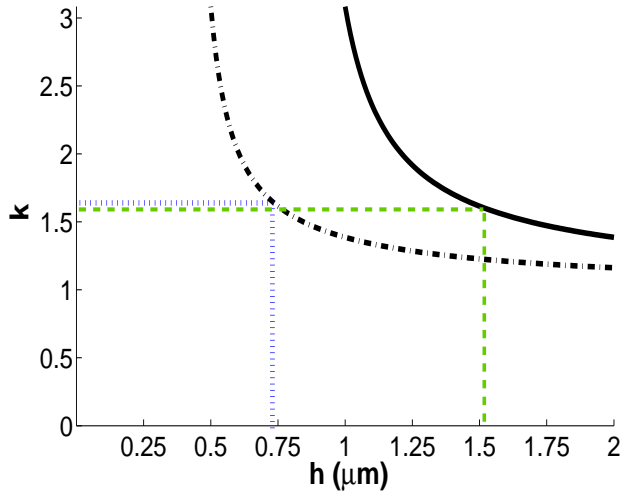}
\caption{ (Color online) $k$ vs $h$ for  $d_1$ (dot-dashed line) and $d_2$ (solid line).  Straight lines show the mean vertical position of  $d_1$ particles (dotted  line) and $d_2 $ particles (dashed line), at $T=20^\text{o}$C. }
\label{k}
\end{figure}

Figure \ref{k} shows the correction factor for the viscosity $k$ as function of $h$, given by Eq.~\eqref{eqk}, for particles of diameters $d_1$ (dot-dashed line) and $d_2$ (solid line). From the values we found for the slopes in Fig. \ref{graf} and using Eq.~\eqref{msdk}, we were able to determine the values of $k$ for each of the studied particle sizes and temperatures. The dotted line and the dashed line in Fig. \ref{k} correspond, respectively, to the values of $k$ found for $d_1$ and $d_2$ at $T=20^o C$, and the associated values of $h$, are determined graphically from the curves for $k$ vs $h$.  
\\

The latex particles sank to the bottom and Fig. \ref{k} is consistent with this behavior, since it implies that the mean vertical position above the bottom surface is around $0.73\mu$m for the smallest particle ($d_1$) and $1.52\mu$m for the largest particle ($d_2$).

\section{Summary}

With a very simple scheme we have observed Brownian motion for 1 and 2$\mu$m-diameter latex spheres, avoiding electrostatic and hydrodynamical interactions among them, using a ionic medium and a very low concentration. The linear behavior of the mean square displacement as function of time was shown for the diffusive regime, in which the sampling time is larger than the relaxation time of the system.\\

Calculations of $N_A$ obtained from a thorough statistical analysis and according to Langevin and Einstein, are very different from standard values. The reason is that the  assumption of an unbounded system was not fulfilled in this case, where the particles sink. This is the most typical situation in systems involving motion of microparticles. The consequence of the sedimentation is that the drag force cannot be assumed to be the Stokes force, as Langevin did.\\

We rather employed the Faxen's correction to the drag, which is a function of the ratio between the diameter of the spheres and the average  separation distance from the  wall. This provides a way to measure $N_A$ in a better approximation or, alternatively, to calculate the average vertical position of the Brownian particle if a standard value of $N_A$ is assumed,  either of both with the same simple setup.\\

Studying Brownian motion can be made in a relatively simple way in an undergraduate laboratory, but it is important to keep in mind that its theory has many subtle issues that could lead to in an improper interpretation of results if they are not considered. When analyzing Brownian systems one must never forget to guarantee the sample to be in thermal equilibrium, to avoid any interaction among particles and to consider boundary conditions of the system, where gravity may play an important role.\\

\begin{acknowledgments}
 We are grateful to Ren\'e Palacios for providing us with useful information on the choice and preparation of the immersion media of the samples. This work was supported by DGAPA-UNAM grant PAPIIT IN100110 and CONACYT M\'exico, grant 132527.
\end{acknowledgments}


\begin{thebibliography}{20}

\bibitem{EINSTEIN} Albert Einstein,  \textit{Investigations on the theory of the Brownian movement}, reprint  (Dover, New York, 1956).
 
\bibitem{SEGRE}  Emilio Segr\`{e}, \textit{From x-rays to quarks}, (Freeman, USA, 1980).  
 
\bibitem{PERRIN} Jean Perrin, \textit{Brownian movement and molecular reality}, translation by F. Soddy (Taylor and Francis, London, 1910).
 
\bibitem{LANGEVIN} L. Don and A. Gythiel,  ``Paul Langevin's 1908 paper: On the Theory of Brownian Motion,'' Am. J. Phys. \textbf{65} (11),  1079-1081  (1997)
 
\bibitem{ATOMS} Jean Perrin, \textit{Atoms}, translation  by D. Hammick  (Constable \& Company LTD, London,  1916).  
 
\bibitem{HARVARD} R. Newburgha, J. Peidle and W. Ruecknerc,  ``Einstein, Perrin, and the reality of atoms: 1905 revisited,'' Am. J. Phys. \textbf{74} (6), 478-481 (2006)     

\bibitem{EJP}  R. Salmon,  C. Robbins and K. Forinash,  ``Brownian motion using video capture'' Eur. J. Phys.  \textbf{23},  249–253 (2002)

\bibitem{BM}  H. Kruglak,  ``Brownian motion - a laboratory experiment''  Phys. Educ. \textbf{23}, 306-309 (1988)

\bibitem{OPTMANIP}  H. Felgner, O. Muller, and M. Schliwa,  ``Calibration of light forces in optical tweezers'' Appl. Opt. \textbf{34} (6), 977-982 (1995)

\bibitem{OPTRAP}  Karen C. Vermeulen, Gijs J. L. Wuite, Ger J. M. Stienen, and Christoph F. Schmidt  ``Optical trap stiffness in the presence and absence of spherical aberrations'' Appl. Opt.\textbf{45} (8), 1812-1819 (2006)  

\bibitem{FAXCOR} J. Leach, H. Mushfique, S. Keen, R. Di Leonardo, G. Ruocco, J. M. Cooper, and M. J. Padgett, ``Comparison of Fax\'en’s correction for a microsphere translating or rotating near a surface'', Phys. Rev. E \textbf{79}, 026301-(1-4) (2009)
  
\bibitem{VANKAMPEN} N. Van Kampen, \textit{Stochastic processes in Physics and Chemistry}, 3rd ed. (Elsevier, Hungary, 2008) 

\bibitem{FAXEN} J. Happel and H. Brenner, \textit{Low Reynolds Number Hydrodynamics}, 2dn ed. (Martinus Nijhoff, Netherlands, 1986) pp. 327.

\bibitem{FAXCOR2} A. Pralle, E. L. Florin, E. H. K. Stelzer, J. K. H. Hörber, ``Local viscosity probed by photonic force microscopy'', Appl. Phys. A: Mater. Sci. Process. \textbf{66}, S71 (͑1998͒)

\bibitem{ByW} M. Born and E. Wolf,  \textit{Principles of optics}, 7th ed. (Cambridge University Press, United Kingdom, 1999).

\bibitem{PART} Duke Scientific Corporation, California, \textit{Bulletin 91M} (2004). 
   
\bibitem{INTERACCIONES2} M. Polin, Y. Roichman, and D. G. Grier, ``Autocalibrated colloidal interactions measurements extended optical traps'', Phys. Rev. E \textbf{77}, 051401-(1-7) (2008)

\bibitem{INTERACCIONES} M. Carbajal, F. Castillo, and J. Arauz, ``Static properties of confined colloidal suspensions'' Phys. Rev. E \textbf{ 53} (4), 3745-3749 (1996)
 
   
\bibitem{SD} A short explanation on Standard Error can be read in the Wolfram Math World web page: $\langle$http://mathworld.wolfram.com/StandardError.html$\rangle$
   
\bibitem{AVOGADRO} Avogadro constant value according with the NIST page: $\langle$http://physics.nist.gov/cgi-bin/cuu/Value?na$\rangle$ 


    
    
  \end{thebibliography}

\end{document}